# Functional near-infrared spectroscopy (fNIRS) and Eye tracking for Cognitive Load classification in a Driving Simulator Using Deep Learning

**Mehshan Ahmed Khan[1], Houshyar Asadi[1], Mohammad Reza Chalak Qazani[2], Chee Peng Lim[1], Saied Nahavandi[3]**

(1) Institute for Intelligent Systems Research and Innovation (IISRI), Deakin University, Geelong, Victoria, Australia: {mehshan.khan, houshyar.asadi, chee.lim}@deakin.edu.au
(2) Facuaty of Computing and Information, Sohar University, Albatinah North Governorate 311, Oman: {mqazani}@ su.edu.om
(3) Swinburne University of Technonology, Hawthorn, Victoria, 3122, Australia: {snahavandi}@swin.edu.au

***Abstract*** *– Motion simulators allow researchers to safely investigate the interaction of drivers with a vehicle. However, many studies that use driving simulator data to predict cognitive load only employ two levels of workload, leaving a gap in research on employing deep learning methodologies to analyze cognitive load, especially in challenging low-light conditions. Often, studies overlook or solely focus on scenarios in bright daylight. To address this gap and understand the correlation between performance and cognitive load, this study employs functional near-infrared spectroscopy (fNIRS) and eye-tracking data, including fixation duration and gaze direction, during simulated driving tasks in low visibility conditions, inducing various mental workloads. The first stage involves the statistical estimation of useful features from fNIRS and eye-tracking data. ANOVA will be applied to the signals to identify significant channels from fNIRS signals. Optimal features from fNIRS, eye-tracking and vehicle dynamics are then combined in one chunk as input to the CNN and LSTM model to predict workload variations. The proposed CNN-LSTM model achieved 99% accuracy with neurological data and 89% with vehicle dynamics to predict cognitive load, indicating potential for real-time assessment of driver mental state and guide designers for the development of safe adaptive systems.*

## Introduction

Driving is a complex activity carried out under tight time constraints, demanding constant cognitive adjustments to the ever-changing driving environment (Shajari, Asadi et al., 2023a, Shajari, Asadi et al., 2023b). While driving a vehicle, drivers are often occupied with many other activities such as using a mobile phone, listening to the radio, or having a conversation with a passenger. Moreover, the integration of advanced in vehicle information systems in modern automobiles introduces the potential for distracted driving scenarios, thereby impacting driving performance (Oviedo-Trespalacios, Haque et al., 2016). Such secondary activities, often unrelated to the primary task of driving, requires extra cognitive processes to ensure that the driver maintains focus on the road and control of the vehicle, even while engaged in other tasks simultaneously. Managing these cognitive load activities while driving requires the allocation of attention and mental resources, often leading to divided attention and potential decreases in driving performance (Chen, Zhao et al., 2024). Thus, understanding how cognitive load affects driving behaviour is necessary for improving road safety and developing effective strategies to address distracted driving risks.

Driving simulators play a vital role in studying human behaviour in complex and challenging environments, which is often impossible in real-world driving situations (Asadi, Lim et al., 2019, Asadi, Mohamed et al., 2015). In these simulated settings, researchers study various aspects of human cognition, decision-making processes, and response patterns under controlled conditions. The n-back task has been commonly used in cognitive psychology, extensively utilized to investigate the intricate workings of working memory in driving (Khan, Asadi et al., 2024). It is known for its ability to elicit activation in the frontoparietal brain regions, widely acknowledged as pivotal for memory function. One of the key strengths of the n-back task lies in its ability to manipulate working memory load incrementally. By adjusting the value of n, researchers can systematically modulate the cognitive demands placed on participants, offering a reliable understanding of working memory engagement. As the n-back levels increase, task performance typically exhibits a discernible decline, accompanied by a perceptible increase in subjective cognitive effort and perceived task difficulty. This relationship between task difficulty, performance decrement, and subjective cognitive effort serves as a crucial psychometric feature of the n-back paradigm.

Researchers have also employed the n-back task as a secondary measure within driving simulation environments to analyse driver performance under conditions not feasible with real vehicles (Huang, Zhang et al., 2024). This approach allows for a detailed examination of cognitive workload, a crucial aspect of driving behaviour. However, assessing cognitive workload is a complex endeavour. Traditionally, methods for evaluating cognitive workload have relied on subjective measures, such as interviews or questionnaire-based approaches, where participants self-report the level of workload experienced during a task (Tingting, Xun et al., 2024).

Several research groups, including Devos et al. (Devos, Gustafson et al., 2020), Wang et al. (Wang, Chardonnet et al., 2021), Y. Zak et al. (Zak, Parmet et al., 2020) and Janczewski et al. (von Janczewski, Kraus et al., 2022), have contributed to the assessment of cognitive workload using subjective methods, primarily through self-assessment questionnaires like the NASA-TLX (National Aeronautics and Space Administration Task Load Index), MCH (Modified Cooper-Harper Scale), and SWAT (Subjective Workload Assessment Test). These questionnaires capture various metrics related to task performance, including mental, physical, and temporal demand, as well as effort, pressure, concentration, and frustration.

In contrast to subjective questionnaire-based methods, evaluating cognitive workload based on physiological signals offers an opportunity for objective and real-time assessment (Khan, Asadi et al., 2023). To enhance efficacy and efficiency, researchers have explored various physiological measures, including Electroencephalography (EEG), Electrocardiography (ECG), functional Near-Infrared Spectroscopy (fNIRS), Eye Tracking, and Heart Rate Variability (HRV). These measures offer a window into the brain's activity, cardiovascular responses, ocular movements, and autonomic nervous system functioning during cognitive tasks. For instance, studies by Ivan et al. (Kesedžić, Šarlija et al., 2020), Kim et al. (Kim, Ryu et al., 2022) and Unni et al. (Unni, Kretzmeyer et al., 2018) have utilized physiological measures to evaluate task performance, employing machine learning (ML) and deep learning (DL) techniques to analyze these signals. By leveraging advanced computational methods, they aim to uncover meaningful patterns and correlations between physiological responses and cognitive workload. This objective approach is considered superior to subjective measures, providing a direct and quantifiable insight into the physiological correlates of cognitive workload (Ghandorh, Khan et al., 2021). Despite significant progress in research, it remains unclear whether driving simulator data alone can effectively predict cognitive load across different levels during n-back tasks in night and rainy weather conditions. Although physiological data holds promise for improving the accuracy of cognitive load classification, its impact on predictive models in low visibility conditions is not yet fully understood. Addressing this gap requires a multimodal approach. Firstly, we need to assess the predictive capabilities of driving simulator data independently. This involves analysing how various features extracted from simulator outputs correlate with different levels of cognitive load, determining the standalone robustness of simulator-based predictions.

In this study, we employed a DL approach to analyse cognitive load during the n-back task within a driving simulator environment, coupled with physiological measures. Numerous studies in the field have traditionally relied mainly on manually crafted features, employing ML techniques to predict cognitive load using physiological signals. However, a notable gap exists in the literature, as many of these studies either overlooked the driving conditions altogether or solely focused on bright daylight conditions. In contrast, our study aims to fill this gap by utilizing the full potential of DL methodologies to analyse cognitive load, specifically in low-light conditions, leveraging data collected from driving simulators, encompassing various parameters such as steering wheel movements, braking, linear acceleration and angular velocity. Furthermore, while traditional studies often utilized the n-back task with only two levels of difficulty, we expanded upon this by incorporating three levels of difficulty. This allowed for a more comprehensive understanding of cognitive workload variations across different task complexities. In addition to exploring the ANOVA feature selection method to understand the feature behaviours, we adopted convolutional neural networks (CNNs) and long short-term memory (LSTM) networks to analyse data collected from the driving simulator, as well as from fNIRS and eye-tracking devices, including fixation duration and gaze directions. This approach enabled us to capture and interpret patterns in both behavioural and physiological data. Moreover, our study has also utilized the environmental conditions which are rarely explored in existing literature, such as driving in low-light nighttime conditions and rainy weather. By incorporating these realistic scenarios, we aimed to simulate more diverse and challenging driving experiences, thus enhancing the ecological validity of our findings.

The subsequent sections of this paper delve into various aspects of the conducted research. In the "Materials and Methods" section, we detail the equipment used for data collection and the experimental procedures followed. A thorough overview of the proposed methodology and its characteristics is presented in the "Data Analysis" section. Within the "Results and Discussion" section, we analyse cognitive load using DL methods. Finally, the "Conclusion" section summarizes the findings and highlight the study's significance.

# Materials and methods

An experiment, using a driving simulator, was designed, encompassing four distinct cognitive load conditions. The conditions entail three different levels of cognitive load conditions: 1) baseline, where no external secondary task was introduced; 2) a lower cognitive load task involving the 0-back task; 3) a medium cognitive load task employing the 1-back paradigm; and 4) a higher cognitive load task challenging participants with the 2-back task. A cohort comprising 10 healthy adults, comprising 9 males and 1 female, was selected through a combination of campus and online recruitment strategies. The participants were stipulated to be regular drivers, engaging in driving activities at least several times per month.

## Driving simulator

Data for the driving simulation was meticulously collected using the "Next Level Racing Motion Platform V3," mounted on a Traction Plus Platform. The simulator setup includes three expansive 32-inch monitors providing a broad horizontal field of view, a Thrustmaster T300 steering wheel, and pedals for control and responsiveness. Configured to replicate the experience of riding in a passenger SUV, this setup ensures an authentic and immersive simulation. Euro Truck Simulator 2 (ETS2) software was utilized for intricate driving scenarios. In this case, the focus was on replicating the challenges of nighttime driving at 1 am, accompanied by simulated rainfall, as depicted in Figure 1. These scenarios not only assess driving skills but also examine how individuals respond to adverse weather conditions and low visibility. Additionally, we have also recorded the car speed, angular velocity X, angular velocity Y, angular velocity Z, linear acceleration X, linear acceleration Y, linear acceleration Z, steering wheel angle, throttle response, and braking response using the ETS2 Software Development Kit (SDK), which will later be used as features to predict different levels of cognitive load.

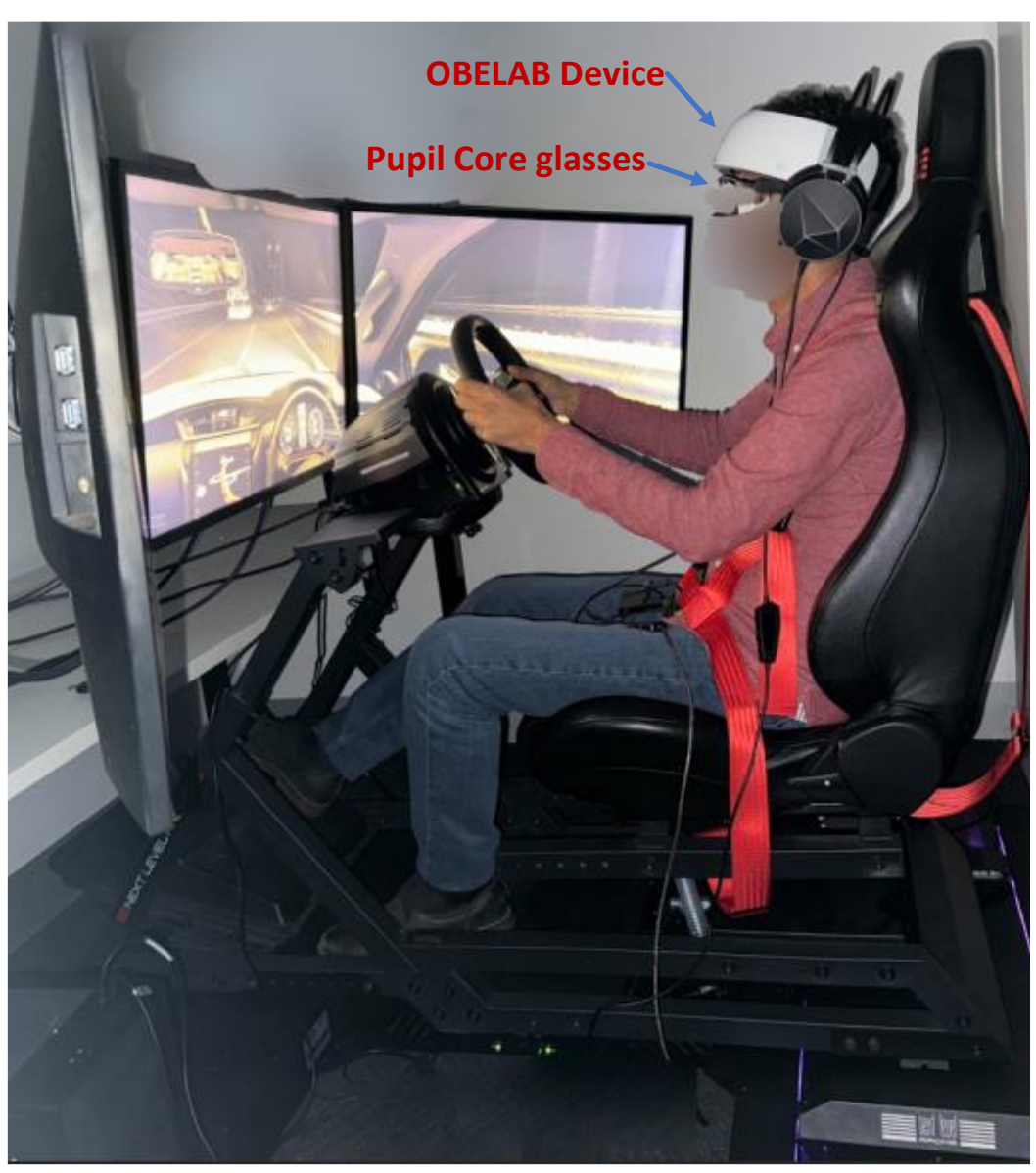

**Figure 1: Analysing cognitive load during simulated nighttime driving.**

## n-back task

To incorporate a manipulation of working memory load into the driving task, we employed an auditory version of the n-back task. This manipulation involved combining a digit-span task with the n-back task to create three distinct n-back levels (n = 0, 1, and 2), with 0-back being the easiest and 2-back being the most challenging. During the task, participants were presented with spoken digits ranging from 0 to 9 in a male voice. Their objective was to compare the currently presented digit with the one that occurred n steps back and respond accordingly by pressing either the red or green buttons on the steering wheel, as illustrated in Figure 2.

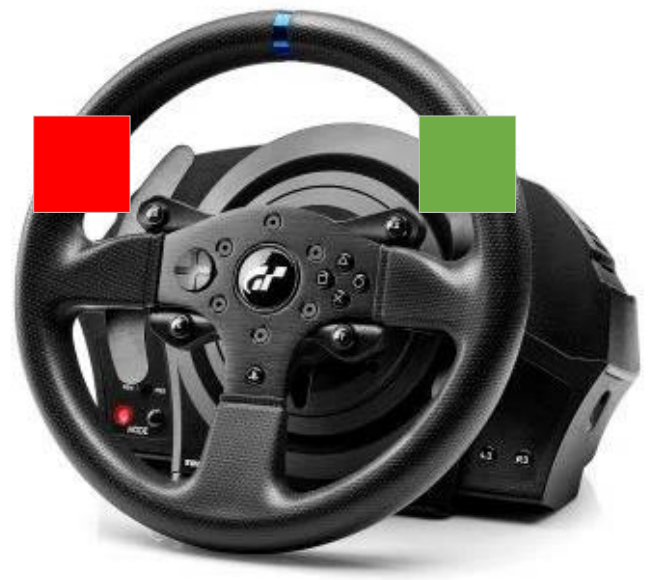

**Figure 2: Steering responses in the n-back task.**

Specifically, participants pressed the green button, located on the right side of the steering wheel, when the response matched the digit presented n trials earlier, and the red button when there was no match. This setup ensured continuous engagement with the task throughout the driving simulation. Auditory stimuli were presented for 1 second, allowing participants a maximum of 2.5 seconds to provide their response before the next digit appeared. They were instructed to respond as quickly as possible to maintain task efficiency.

To implement the n-back task and record responses, we utilized Psychopy (Peirce, Gray et al., 2019) software. The design of the n-back tasks ensured that participants heard both the digit sounds and their surrounding environment simultaneously, adding an additional layer of cognitive complexity during the driving simulation.

## Physiological measures

We have gathered driving data alongside fNIRS and eye-tracking data to predict cognitive load of different levels. To estimate the concentration of oxyhaemoglobin (HbO2) and deoxyhaemoglobin (HbR) from the prefrontal cortex, we will use a portable fNIRS device called NIRSIT (developed by OBELAB in Seoul, South Korea). This device offers a 48-channel configuration with two wavelengths of 780 nm and 850 nm. The 48 channels correspond to specific regions within the prefrontal cortex, as illustrated in Figure 3. To obtain the desired measurements, the fNIRS data collected by the device is processed using the modified Beer-Lambert Law (Baker, Parthasarathy et al., 2014). This law is employed to convert the raw fNIRS signals into meaningful information about the concentration of oxyhaemoglobin. The NIRSIT device incorporates source-detector pairs, which consist of 24 dual-wavelength laser diodes and 32 photodetectors. These pairs are spaced at 1.5 cm from each other, enabling the acquisition of data from multiple distances.

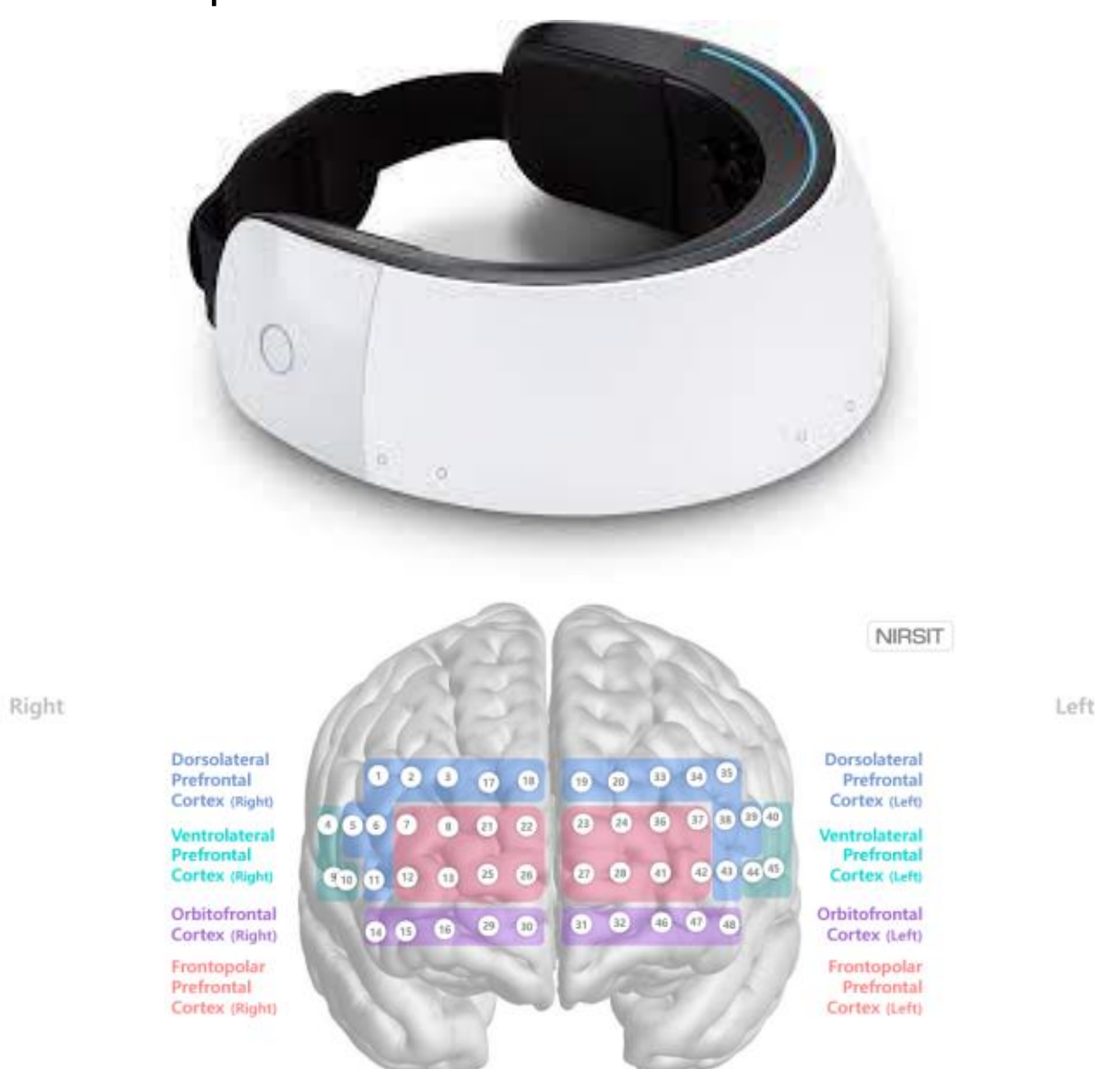

**Figure 3: Spatial Distribution of NIRSIT 48 Channels in the Prefrontal Cortex. (Image taken from OBELAB. NIRSIT Channel Information)**

In addition to the physiological signals obtained through fNIRS, we have also acquired eye-tracking data using Pupil Core glasses as shown in Figure 4. Eye-tracking technology allows for the precise tracking and analysis of participants' eye movements, fixations duration and gaze patterns during the cognitive tasks.

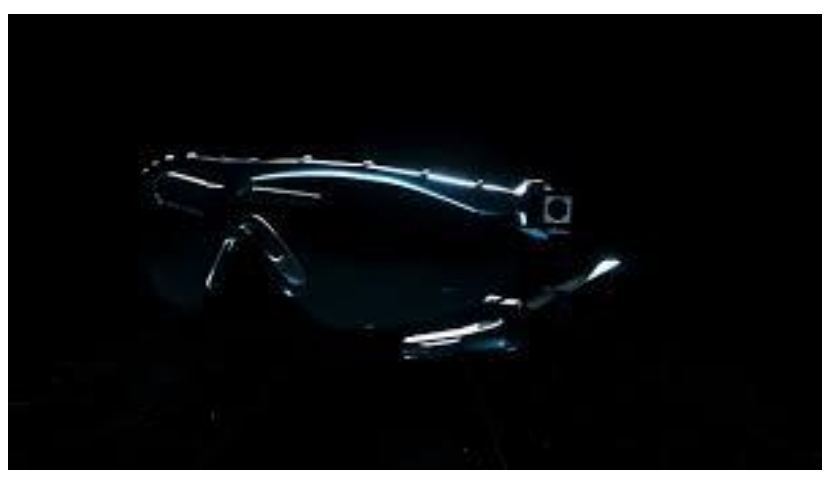

**Figure 4: Pupil Core glasses.**

Eye-tracking devices are useful for directly measuring eye movements in response to different cognitive tasks (Chen, Chen et al., 2023). These devices provide valuable information into the working principle of the human mind, as eye movements provide an indication of where an individual's attention is focused, the level of engagement, and the cognitive workload (Frischen, Bayliss et al., 2007, Scott and Hand, 2016).

## Experimental procedure

Upon participant's arrival, their eligibility was verified, and informed consent was obtained. To ensure standardized instructions, a prerecorded video featuring a male voice demonstration of the driving simulator controls and the n-back task was played for all participants. Subsequently, any queries pertaining to the n-back task and the operation of the driving simulator were addressed comprehensively. Prior to engaging in the experimental drives, participants underwent a practice session for the n-back task to ensure familiarity and proficiency. This practice session was conducted separately from the driving simulator, allowing participants to focus solely on mastering the cognitive task. Once participants demonstrated a thorough understanding of the n-back task, they proceeded to a practice drive in the simulator.

The driving simulator replicated a freeway environment, where participants navigated through virtual highways at a speed limit of 113 km/hr. To enhance realism and challenge, the simulation was set at nighttime, precisely at 1 am, with simulated rain adding an additional layer of complexity. Throughout the experiment, windshield wipers were activated, ensuring clear visibility despite the simulated rainfall. By providing standardized instructions and practice sessions, we aimed to minimize variability among participants and ensure a consistent baseline for assessment. The integration of realistic driving conditions in the simulator allowed for the evaluation of cognitive load in a controlled yet dynamic environment, highlighting the cognitive requirements linked to driving activities under challenging conditions.

# Data analysis

This section details about the CNN-LSTM structure proposed within the paper. The core architecture of the model is depicted in Figure 5, showcasing the integration of two neural networks: CNN and LSTM.

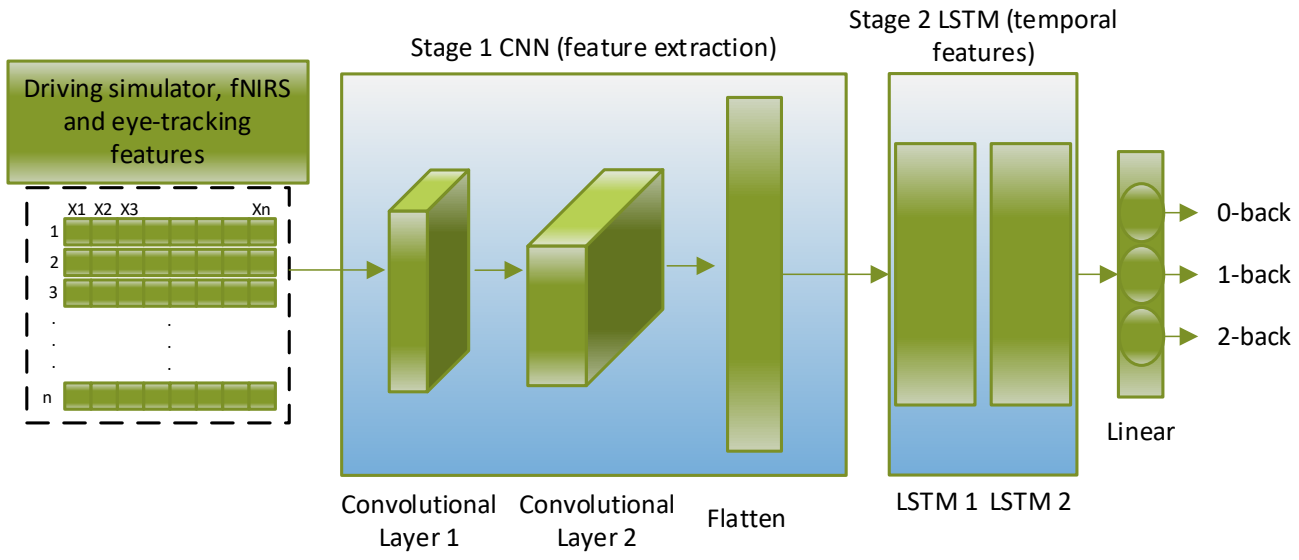

**Figure 5: Proposed CNN and LSTM architecture for analyzing cognitive load in simulated environments.**

Our approach involves the development of a prediction model that utilizes the capabilities of both CNN and LSTM architectures (Khan, Asadi et al., 2024). The proposed CNN-LSTM model represents a novel approach to predictive modelling, leveraging the strengths of CNN and LSTM networks to enhance performance and accuracy.

## Convolutional Neural Networks (CNNs)

In our research, we employed 2 layers of 1-dimensional CNN (1D-CNN) to extract effective and representative features from 1D time-series sequence data. This involved utilizing 1D convolution operations with multiple filters to capture relevant temporal patterns. Given that our data consists of one-dimensional signals, such as time-series data, we utilized Convolution 1D layers, pooling 1D layers, and a fully connected layer to construct our CNN architecture. Let $X$ be the input time-series data, where $X = \{x_1, x_2, x_3, \dots, x_n\}$ represents the sequential time instants. In the convolutional layer, feature maps $F$ are generated by convolving the input data $X$ with learnable filters $W$ and applying a bias term $b$, followed by an activation function σ:

$$F_i = \sigma(\mathrm{W} * \mathrm{X} + \mathrm{b}) \quad (1)$$

In Eq. 1. $F_i$ represents the $i^{th}$ feature and $\mathrm{W} * \mathrm{X}$ represents the convolution operation.

## Long Short-Term Memory (LSTMs)

In our approach, the output of the CNN serves as the input to the LSTM (Boulila, Ghandorh et al., 2021), allowing the LSTM to capture and integrate temporal information extracted by the CNN. Unlike traditional Recurrent Neural Networks (RNNs), LSTMs address issues such as vanishing and exploding gradients that impede learning long-term dependencies in sequential data (Wang, Peng et al., 2020). This is achieved through specialized memory cells and gating mechanisms that facilitate selective retention and discarding of information over extended sequences. LSTMs possess internal memory units that enable them to retain and utilize information from previous time steps, enabling effective modeling of cognitive load dynamics. This capability is particularly relevant for driving simulator, fNIRS, and eye-tracking data, which exhibit dependencies on previous observations. The equations governing the behavior of CNN-LSTM models can be described as follows:

$$i_t = \sigma(W_{xi} * X_t + W_{hi} * H_{t-1} + W_{ci} \circ C_{t-1} + b_i), \quad (2)$$

$$f_t = \sigma(W_{xf} * X_t + W_{hf} * H_{t-1} + W_{cf} \circ C_{t-1} + b_f), \quad (3)$$

$$o_t = \sigma(W_{xo} * X_t + W_{ho} * H_{t-1} + W_{co} \circ C_{t-1} + b_o), \quad (4)$$

$$\tilde{C}_t = \tanh(W_{xc} * X_t + W_{hc} * H_{t-1} + b_c) \quad (5)$$

$$C_t = f_t \circ C_{t-1} + i_t \circ \tilde{C}_t \quad (6)$$

$$H_t = o_t * \tanh(C_t) \quad (7)$$

In Eq. 2, Eq. 3, Eq. 4, Eq. 5, Eq. 6 and Eq. 7, '$*$' represents convolution, and '$\circ$' represents the Hadamard product. Cell states are denoted as $C_1, \dots, C_t$ and hidden states as $H_1, \dots, H_t$. The input gate is denoted as $i_t$, the forget gate as $f_t$, and the output as $o_t$. The sigmoid activation function, $\sigma$ adjusts the gate output between 0 and 1. $W_{xi}$, $W_{hi}$, $W_{ci}$, $W_{xf}$, $W_{hf}$, $W_{cf}$, $W_{xo}$, $W_{ho}$, $W_{co}$, $W_{xc}$, $W_{hc}$ are convolution kernels. $b_i$, $b_c$, $b_f$ and $b_o$ are the bias terms.

## Proposed CNN-LSTM based architecture

The model architecture incorporates both CNN layers, which excels at extracting spatial features from input data, enabling robust feature representation through convolutional operations. (Khan, Khan et al., 2020). On the other hand, the LSTM network specializes in capturing temporal dependencies within sequential data, facilitating the modelling of long-range dependencies and dynamic patterns over time. (Boucetta, Amrane et al., 2024). This choice is apt for our dataset consisting of driving simulator data and fNIRS, which exhibit both spatial features and time series characteristics. Study conducted by (Pham, Hoang et al., 2018) has emphasized the benefits of combining 1D convolutional layers with LSTM layers, particularly in improving accuracy, compared to models relying solely on LSTM layers. Hence, we adopted a hybrid 1D CNN and LSTM architecture in building our system to efficiently train the features, harnessing the strengths of both architectures.

In the workflow of our CNN–LSTM model, depicted in Figure 5, the dataset undergoes preprocessing before being divided into training and testing sets. Subsequently, data normalization is carried out to ensure uniformity and facilitate effective training. The normalized data is then inputted into the model. The model commences with two CNN layers, with kernel sizes of 3 and output channels of 16 and 32, respectively. These layers are followed by rectified linear unit (ReLU) activation functions to capture spatial features from the input data. The LSTM layer, with an input size of 32, hidden size of 64, and two layers, captures temporal dependencies in sequential data. Configured with a batch-first arrangement, this LSTM network comprehends temporal dynamics in cognitive load over time, which is crucial for understanding fluctuations during driving tasks. Subsequently, the model flattens the LSTM output and passes it through two fully connected layers with sizes of 64 and 128, respectively, followed by a ReLU activation function. This process aligns these features with a final output size of 3, representing the three cognitive load levels. The network's size is kept modest to enable real-time prediction of cognitive load.

# Results and discussion

To begin the assessment of cognitive load using driving data, a preprocessing step was implemented to ensure uniformity and suitability for analysis. This involved employing the standard scalar function on both the driving simulator data and physiological measures. The standard scalar preprocessing technique normalizes and scales the data to establish a mean of zero and a standard deviation of one, thereby ensuring all features are on a standardized scale. This normalization facilitates the processing of data by ML and DL algorithms, enhancing efficiency and accuracy in analysis.

## Feature analysis of driving data

In order to analyse the cognitive processes across three distinct levels, a comprehensive analysis was conducted on various features extracted from the driving simulator. Employing ANOVA, an analytical tool widely used for comparing means across multiple groups (Lakshmi and Maheswaran, 2024), we analysed the driving parameters to uncover significant insights. The ANOVA feature analysis revealed that the car speed emerging as the most significant factor in discerning cognitive load levels across the spectrum. Figure 6 illustrates impact of car speed on identifying cognitive load distinctions among different levels. The velocity of the car stands out as a significant feature, potentially attributed to its profound impact on cognitive processing, particularly in scenarios where higher speeds demand greater cognitive resources for maintaining control. This phenomenon could be indicative of a strategic adaptation wherein individuals consciously moderate their driving speed to accommodate the additional cognitive load imposed by the task complexity. Such adjustments likely reflect a proactive effort to enhance cognitive bandwidth available for task performance, potentially mitigating the risk of compromised control associated with heightened cognitive demands at higher speeds.

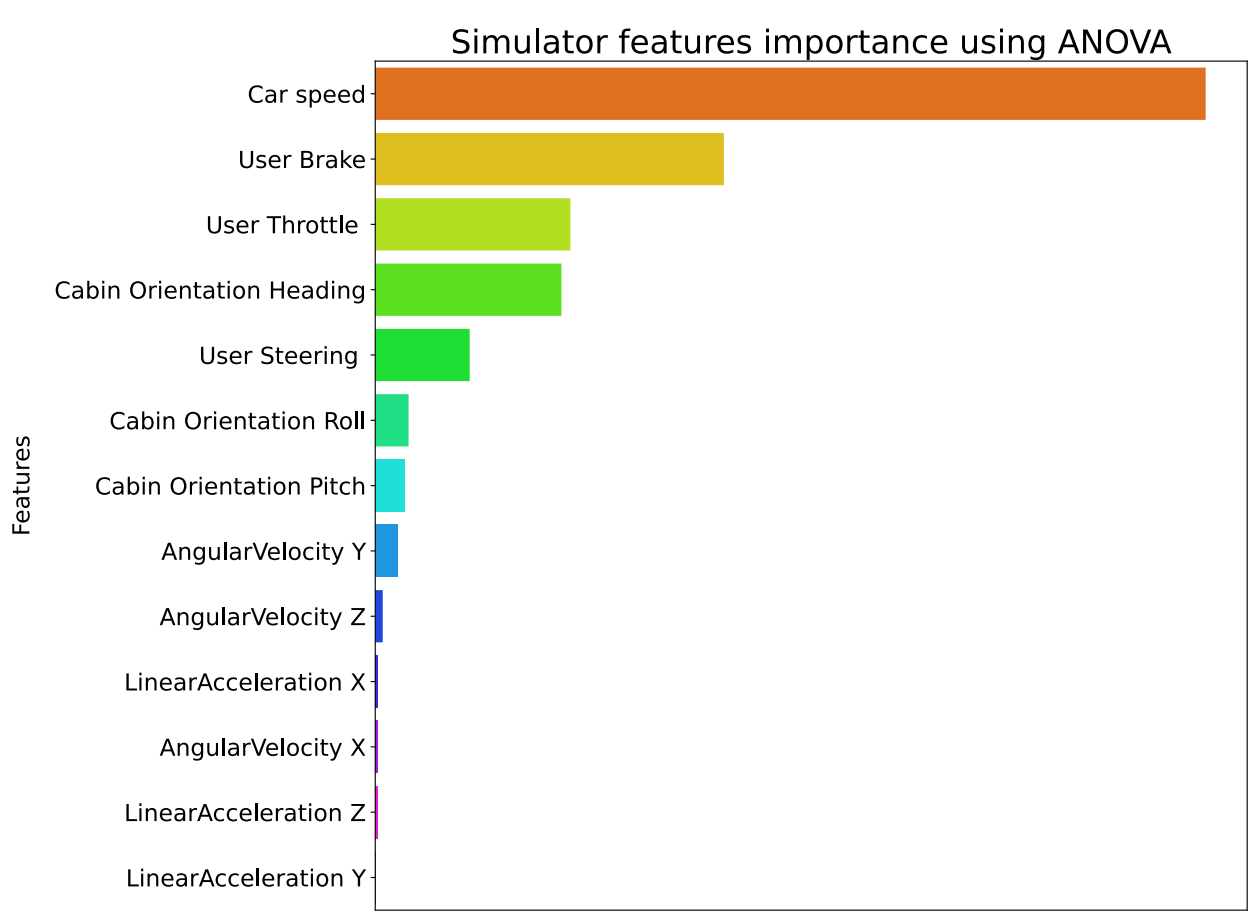

**Figure 6: Visualization depicting the simulator feature importance plot generated through ANOVA analysis.**

Additionally, aside from car speed, several other factors such as brake usage, steering behaviour, angular velocity, linear acceleration, and cabin orientation were examined. However, their significance appears to be relatively minor, potentially because they rely on the driver's experience. Even though these variables may influence driving performance, car speed appears to have a greater influence overall, particularly when it comes to cognitive workload. Unlike car speed, which directly impacts cognitive load due to its implications for control and decision-making, these other factors may depend more on individual driver skill and familiarity with driving tasks.

## Feature analysis of physiological

In addition to the driving data, we have utilized fNIRS and eye-tracking technology to analyse cognitive load. These sensors were synchronized with varying levels of workload to capture real-time cognitive responses. Our fNIRS device utilizes 38 channels for practical use but encompasses 202 channels to ensure robustness against noise interference. In instances where data from certain channels becomes tainted by noise, backup channels seamlessly replace them to maintain data integrity. For our analysis, we harnessed the entirety of these 202 channels. To extract meaningful features from the fNIRS data, we employed ANOVA across all channels. This statistical approach allowed us to discern the channels that yielded significant influence in identifying cognitive load. For our ML/DL analyses of cognitive load levels, we chose the top 20 features extracted from fNIRS, as depicted in Figure 7. Analysing these features through ANOVA, we found that HbO2 features have a more pronounced influence compared to HbR features. Statistically, HbO2 features show a greater impact on assessing different levels of cognitive workload than their HbR counterparts.

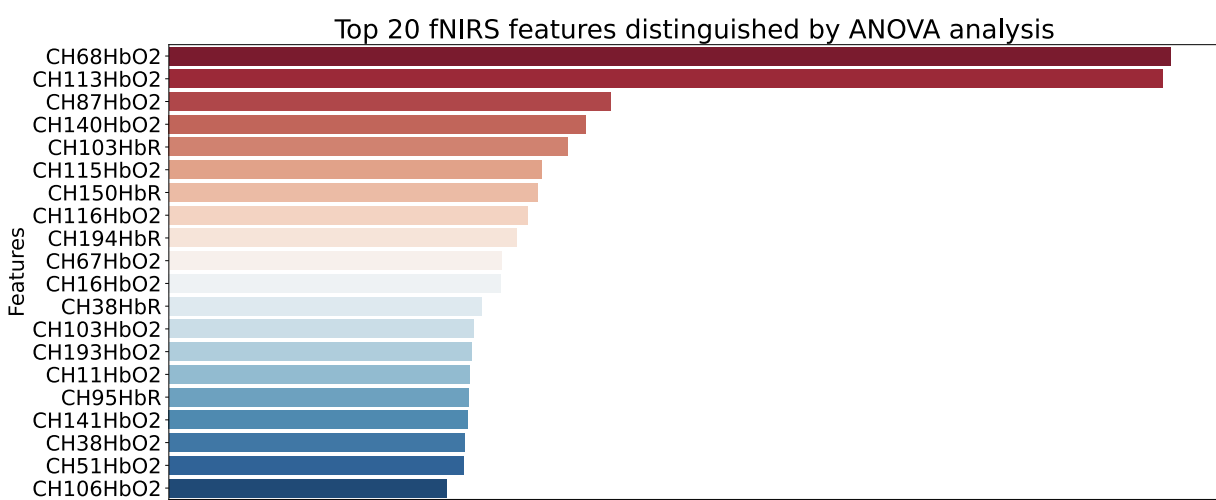

**Figure 7: fNIRS features for cognitive load analysis.**

## Fusion of driving simulator, fNIRS and eye-tracking

An essential consideration during this integration process was the disparate sampling rates of the simulator data compared to that of fNIRS and eye-tracking data. Given the higher sampling rate of simulator data, a down-sampling (Wu, Ye et al., 2024) procedure was implemented to ensure compatibility with the sampling rates of fNIRS and eye-tracking data. This down-sampling approach was executed to maintain the temporal resolution across all data modalities, resulting in feature vectors of uniform length across the dataset. In our merged dataset, we have incorporated various data streams to provide a thorough understanding of cognitive load dynamics. This integration encompasses driving simulator metrics like car speed, angular velocities in X, Y, Z directions, linear accelerations in X, Y, Z directions, steering wheel angle, throttle response, and braking response. Alongside these simulator parameters, we have included the top twenty features extracted from fNIRS data, capturing both HbO2 and HbR responses. Furthermore, our dataset integrates eye-tracking data, including fixation duration and gaze direction. Figure 8 presents a correlation map displaying the relationships among all the features.

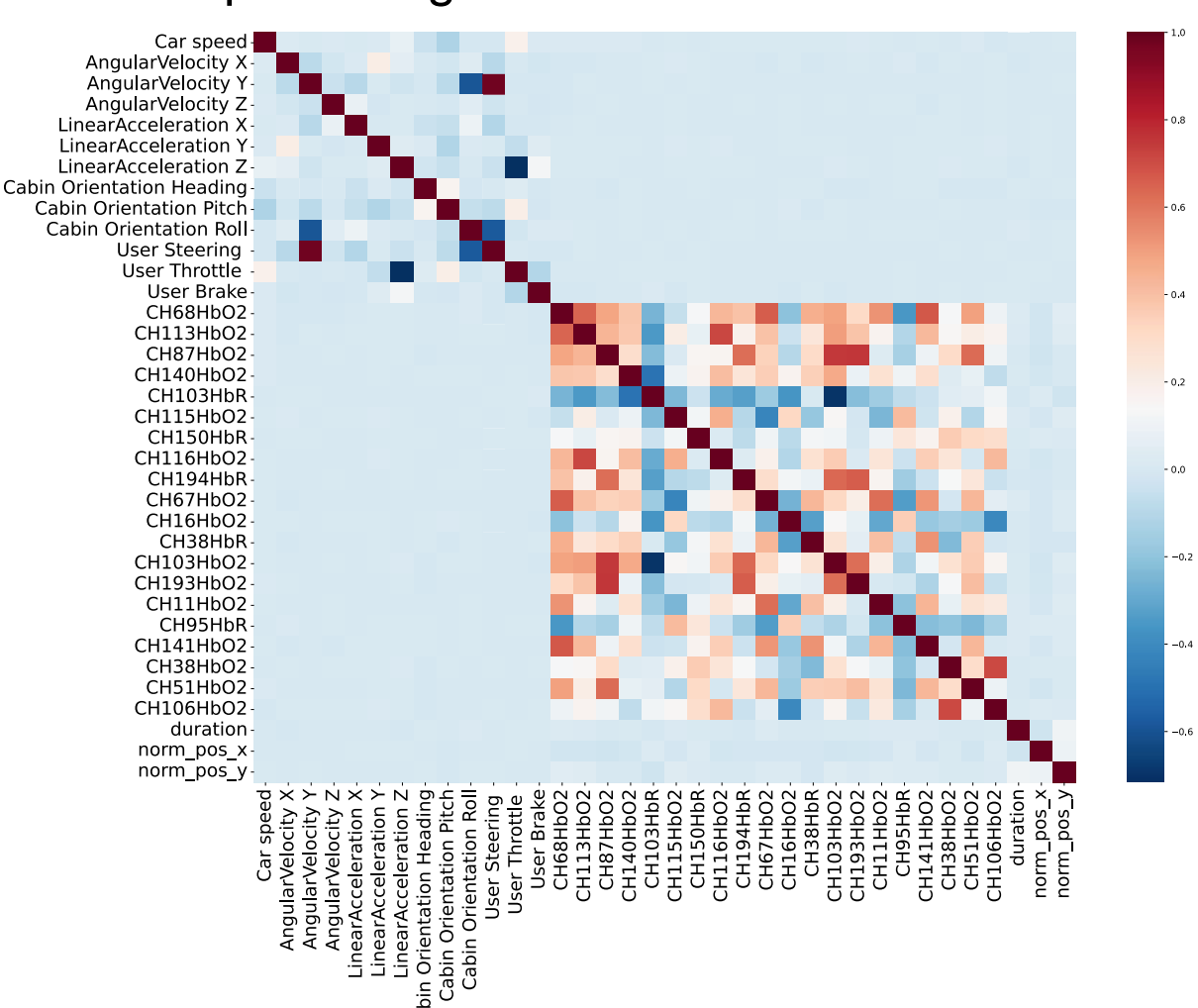

**Figure 8: Correlating driving dynamics with physiological signals.**

The visualization highlights a significant correlation between steering behaviour and cabin orientation with angular velocity. This implies that changes in steering input and the orientation of the vehicle's cabin are closely linked to variations in angular velocity.

## ML/DL classification results

The study utilizes Python, in conjunction with the PyTorch framework (Paszke, Gross et al., 2017), to enable a strong implementation and analysis of models. Additionally, the computational setup includes an Intel Core i9 12900K CPU, 32 GB of RAM, and an NVIDIA RTX 3090ti GPU for model training and evaluation.

In our DL model training, we employed the Adam optimizer with a specified learning rate of 0.001. The Adam optimizer is well-suited for training neural networks, as it efficiently adjusts the learning rate during training to adapt to the varying gradients of the loss function. Moreover, we incorporated the cross-entropy loss function into our model architecture. Cross-entropy loss is commonly used in classification tasks and is particularly effective when training models to output probability distributions over multiple classes. Our training process spanned a total of 1000 epochs, allowing us to thoroughly evaluate the model's progression over time.

To evaluate the efficacy of the proposed model, we conducted an analysis of the driving simulator data. Our proposed CNN-LSTM model demonstrated superior performance, achieving an accuracy of 89.60%. Table 1 presents a comprehensive comparison of the proposed CNN-LSTM model with other ML classifiers using driving simulator data. Furthermore, we conducted comparative analyses by comparing the results of our proposed CNN-LSTM model against those of classical ML classifiers. Notably, our proposed CNN-LSTM model outperformed classical ML classifiers in terms of F1-score, precision, recall, and Area Under the Curve (AUC). This superior performance across a spectrum of evaluation criteria reaffirms the efficacy and superiority of our novel approach in cognitive load assessment. However, it's essential to note that certain classical ML classifiers, such as decision trees, exhibited comparable performance to our proposed model. Decision trees yielded an accuracy of 87.26% and demonstrated similarity with our model across other metrics. Conversely, Naïve Bayes and Nearest Centroid classifiers yielded suboptimal performance, with accuracy of around 35% and AUC values approximately close to 50%.

**Table 1: CNN-LSTM's comparative Performance with driving simulator data only.**

| Model | Accuracy | F1-score | Precision | Recall | AUC |
|---|---|---|---|---|---|
| **Naïve Bayes** | 0.3477 | 0.2587 | 0.3698 | 0.3477 | 0.5069 |
| Nearest centroid | 0.3546 | 0.3483 | 0.3526 | 0.3546 | 0.5286 |
| k-NN | 0.5761 | 0.5747 | 0.5812 | 0.576 | 0.6659 |
| Decision trees | 0.8726 | 0.8726 | 0.8726 | 0.8726 | 0.9056 |

| **Proposed CNN-LSTM** | **0.8960** | **0.896** | **0.896** | **0.896** | **0.9218** |
|---|---|---|---|---|---|

To enhance the performance of the proposed CNN-LSTM architecture, we conducted a comprehensive evaluation by merging driving simulator data with fNIRS and eye-tracking features, as illustrated in Figure 8. This integration resulted in a significant performance increase of approximately 10% with the CNN-LSTM model and decision trees. Furthermore, the confusion matrix depicted in Figure 9 illustrates minimal misclassifications observed in the proposed CNN and LSTM models.

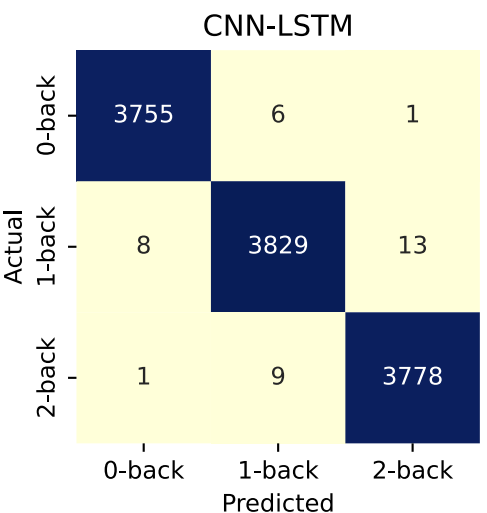

**Figure 9: Confusion matrix depicting the performance of the proposed CNN-LSTM model utilizing integrated driving simulator, fNIRS, and eye-tracking data.**

Moreover, compared to other ML classifiers, we observed an improvement in performance of nearly 20%. Figure 10a, Figure 10b, Figure 10c and Figure 10d illustrate the confusion matrices for Nearest Centroid, k-NN, Naïve Bayes, and Decision Trees respectively.

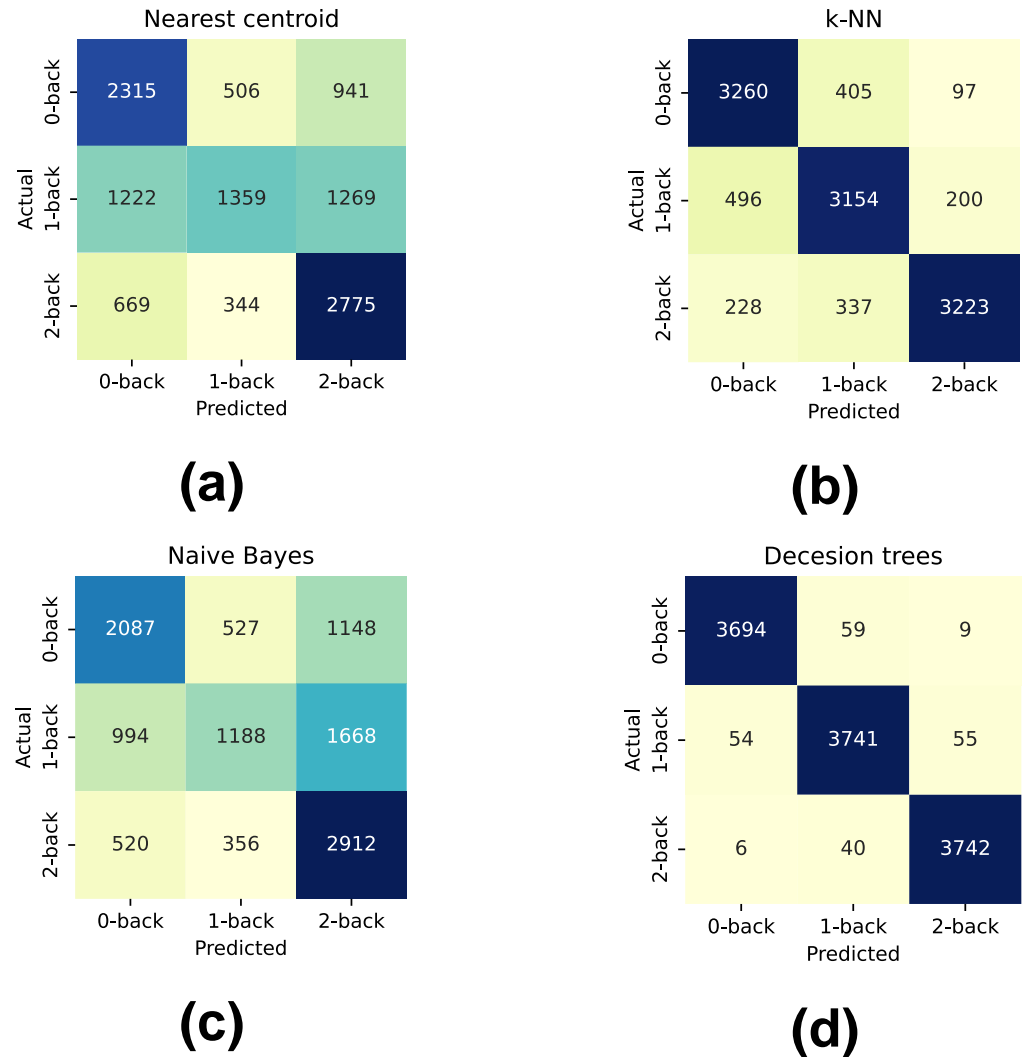

**Figure 10: Confusion matrices illustrating the performance of (a) Nearest Centroid, (b) k-NN, (c) Naïve Bayes, and (d) Decision Trees using integrated driving simulator, fNIRS, and eye-tracking data.**

Particularly noteworthy is the superior performance of the proposed CNN-LSTM framework over other classifiers, demonstrating its effectiveness in utilizing multimodal data for predicting cognitive load. Table 2 provides a comprehensive breakdown of key metrics and performance indicators.

**Table 2: Analyzing CNN-LSTM performance against ML classifiers with driving Simulator, fNIRS, and eye-tracking data.**

| **Model** | **Accuracy** | **F1-score** | **Precision** | **Recall** | **AUC** |
|---|---|---|---|---|---|
| **Naïve Bayes** | 0.5427 | 0.5259 | 0.5539 | 0.5427 | 0.7029 |
| Nearest centroid | 0.5657 | 0.5534 | 0.5743 | 0.5657 | 0.7205 |
| k-NN | 0.8453 | 0.8458 | 0.8476 | 0.8453 | 0.9137 |
| Decision trees | 0.9804 | 0.9804 | 0.9804 | 0.9804 | 0.9919 |
| **Proposed CNN-LSTM** | **0.9982** | **0.9967** | **0.9967** | **0.9967** | **0.9982** |

The proposed model integrates vehicle dynamics, fNIRS, and eye-tracking data for the assessment of cognitive load, particularly in dynamic scenarios like driving. Our study demonstrates the efficacy of the CNN-LSTM model in accurately predicting cognitive load levels by combining features from different modalities, resulting in a significant performance improvement compared to traditional ML classifiers. Figure 7 illustrates how the model identifies specific fNIRS channels crucial for real-time assessment, capturing neural activity indicative of cognitive load levels. Leveraging only driving simulator data, our model achieved an 89% accuracy in predicting workload across varying cognitive load levels, demonstrating the reliability of this approach. By focusing solely on driving simulator parameters such as car speed, angular velocities, accelerations, steering wheel angle, throttle response, and braking response, our model effectively analyzes cognitive workload dynamics.

However, it's crucial to acknowledge certain limitations of our study. Firstly, the sample size utilized for model training and evaluation may not fully represent the diverse population of drivers, potentially limiting the generalizability of our findings. Additionally, while our integration of multiple data streams enhances predictive accuracy, it also introduces complexity in data preprocessing and feature extraction, which may not be feasible in all practical applications. Furthermore, the use of specific sensors like fNIRS and eye-tracking devices may pose challenges in real-world implementation due to their cost and technical requirements. Despite these limitations, our study holds the potential of multimodal data fusion in cognitive load assessment for future research in this domain. Further investigations could explore additional features and data sources to enhance model robustness and generalizability.

# Conclusion

This study presents a comprehensive approach to evaluate cognitive load levels using a CNN-LSTM

model. The model is capable of predicting cognitive load by analysing both neurological data and vehicle dynamics data, or solely vehicle dynamics data. In addition to utilizing fNIRS and eye-tracking, our model demonstrates the ability to accurately predict cognitive load solely based on vehicle dynamics information. This capability is particularly valuable for the development and implementation of vehicle adaptive systems, which rely on real-time assessments of driver cognitive load to enhance safety and performance. .To explore the relationship between driving conditions and mental workload, we gathered data from a group of 10 experienced drivers. Our experiments took place in a simulated environment designed to mimic nighttime driving in rainy conditions, simulating challenging real-world scenarios. We ensured the continuous engagement in the auditory n-back task to induce mental workload while participants engaged in driving tasks. Integrating the n-back task into our driving simulations, our aim was to observe how drivers cope with increased cognitive demands and adverse weather and low lighting conditions. In contrast to previous studies that mainly focus on binary classifications of cognitive load, our study takes a different approach by incorporating three levels of cognitive load classification.

Through preprocessing and feature analysis, we identified key driving parameters and physiological features crucial for cognitive load assessment. Our findings highlight the significant impact of car speed on discerning cognitive load levels, alongside the prominence of HbO2 features in fNIRS data analysis. The integration of these diverse physiological signals into a CNN-LSTM, demonstrating superior performance in predicting cognitive load compared to traditional ML classifiers such as naïve bayes, decision trees, k-NN and nearest centroid. By the integration of diverse data sources like vehicle dynamics, fNIRS, and eye-tracking metrics, our model achieved an accuracy of 99.82% in predicting cognitive load. Additionally, our research highlights the effectiveness of DL models in predicting cognitive workload based solely on vehicle dynamics, with an accuracy of 89%. This highlights the capacity of DL models in predicting mental workload by observing driving behaviour. Future work should focus on the advancement of DL models for integration into adaptive vehicle systems. These systems could continuously monitor driver cognitive load in real-time and intervene as needed, potentially assuming control from drivers during periods of increased cognitive demand. Such an approach holds promise for enhancing road safety by offering dynamic assistance that responds to drivers' cognitive states, thus reducing the risk of accidents associated with cognitive overload.